\documentclass[twocolumn]{aastex631}

\begin{document}

\title{Shedding light on the origin of the broken misaligned circumtriple disk around GW Ori}

\author[0000-0002-4314-398X]{Jeremy L. Smallwood}
\affiliation{Institute of Astronomy and Astrophysics, Academia Sinica, Taipei 10617, R.O.C.}
\affiliation{Homer L. Dodge Department of Physics and Astronomy, The University of Oklahoma,
Norman, OK 73019, USA}

\author[0000-0002-4636-7348]{Stephen H. Lubow}
\affiliation{Space Telescope Science Institute, Baltimore, MD 21218, USA}

\author[0000-0003-2401-7168]{Rebecca G. Martin}
\affiliation{Nevada Center for Astrophysics, University of Nevada, Las Vegas, 4505 South Maryland Parkway, Las Vegas, NV 89154, USA}
\affiliation{Department of Physics and Astronomy, University of Nevada, Las Vegas,
4505 South Maryland Parkway 
Las Vegas, NV 89154, USA}

\author[0000-0003-0856-679X]{Rebecca Nealon}
\affiliation{Centre for Exoplanets and Habitability, University of Warwick\\
Coventry CV4 7AL, UK\\
}
\affiliation{Department of Physics, University of Warwick\\
Coventry CV4 7AL, UK\\
}



\begin{abstract}
We revisit the origin of the observed misaligned rings in the circumtriple disk around GW Ori. Previous studies appeared to disagree on whether disk breaking is caused by the differential precession driven in the disk by the triple star system. In this letter, we show that the previous studies are in agreement with each other when using the same set of parameters. But for observationally motivated parameters of a typical protoplanetary disk, the disk is unlikely to break due to interactions with the triple star system. We run 3-dimensional hydrodynamical simulations of a circumtriple disk around GW Ori with different disk aspect ratios. For a disk aspect ratio typical of protoplanetary disks, $H/r  \gtrsim 0.05$, the disk does not break. An alternative scenario for the gap’s origin consistent with the expected disk aspect ratio involves the presence of giant circumtriple planets orbiting GW Ori.
\end{abstract}



\section{Introduction} 
\label{sec::intro}

The origin of the misaligned broken disk around GW Ori has sparked conversation in the community. The key question centers on whether the disk is broken due to interactions with the triple star system \citep[e.g.][]{Larwood1997,Nixon2013} or if it is instead influenced by a potential circumtriple planet. Both mechanisms are plausible for inducing disk breaking in protoplanetary disks \cite[e.g.,][]{Facchini2013,Dong2018,Young2023,Rabago2024}. However, in this letter, we present evidence that the misaligned broken disk around GW Ori is unlikely to be caused by interactions with the triple stars alone. Our analysis suggests that alternative explanations, such as the presence of a circumtriple planet, may play a more significant role in driving the observed misaligned disk structure.

\cite{Kraus2020}, \cite{Bi2020}, and \cite{Smallwood2021} all conducted smoothed particle hydrodynamical (SPH) simulations of the misaligned protoplanetary disk around GW Ori to investigate the origin of the large gap observed in the disk. Both \cite{Bi2020} and \cite{Smallwood2021} used the SPH code {\sc phantom} \citep{Price2018}, while \cite{Kraus2020} used sphNG \citep{Benz1990}. The simulation results from \cite{Bi2020} and \cite{Smallwood2021} were consistent with each other, but differed from those of \cite{Kraus2020}. In a media article, \cite{kraus2020media} argued that the discrepancy in the results from \cite{Bi2020} was that they modeled a binary star system rather than a triple star system. However, \cite{Smallwood2021} demonstrated that the inner A-B binary can be modeled as a single star, given its small separation compared to the disk extent. They ran N-body simulations comparing the orbital evolution of the triple star system to that of the binary approximation and found that using the binary approximation does not significantly alter the system’s orbital dynamics for the moderate misalignment observed between the triple star orbit and the disk inclination \citep[see also][]{Lepp2023}. Furthermore, SPH simulations comparing the two and three star system models showed little difference in their evolution. Therefore, the binary approximation is not the cause of the discrepancy in the SPH results between \cite{Bi2020} and \cite{Kraus2020}.

The first version of \cite{kraus2020v1} (version 1) indicated that their SPH simulations used a disk aspect ratio of $H/R=0.05$. \cite{Smallwood2021} used this disk aspect ratio and found no evidence of disk breaking for physically motivated parameters of a protoplanetary disk. In an updated version of \cite{Kraus2020} (version 2), the disk aspect ratio was changed to $H/R=0.02$, however, the simulation results remained unchanged, suggesting that the initial disk aspect ratio was likely a typographical error.  In this paper we will explore the implications of that error.


 A misaligned disk experiences gravitational torque from the triple star system, resulting in differential precession that can cause warping, breaking \citep{Nixon2013a,Facchini2013}, or tearing \citep{Nixon2012a}. The dissipation within the misaligned disk leads to the formation of a warp, which evolves in either the diffusive or the bending wave regime, depending on the disk's thickness and viscosity \citep{Paploizou1983,papaloizou1995,Ogilvie1999,Nixon2016}. In the bending wave regime, where the disk's aspect ratio exceeds the viscosity coefficient ($H/r \gtrsim \alpha$), the warp induced by the binary torque propagates as a pressure wave at a speed of  one half the gas sound speed $c_{\rm s}/2$ \citep{Paploizou1983,papaloizou1995}. In the diffusive regime, where the aspect ratio is smaller than the viscosity ($H/r\lesssim \alpha$), the diffusion coefficient is inversely proportional to the disk viscosity. Protoplanetary disks are generally expected to be in the bending wave regime \citep{Hartmann1998}.

For the disk aspect ratio of $H/R=0.05$, the disk is in the bending wave regime. For  disk aspect ratio $H/R=0.02$, 
the disk is close to the boundary between the bending wave and viscous regimes.

Disks in the bending wave regime break more easily
for smaller disk aspect ratio \cite[e.g.,][]{Larwood1997}. 
This behavior occurs because a cooler disk is in weaker radial communication by pressure for maintaining rigid rotation without breaking. We then expect that the model with $H/R=0.02$ should break more easily than the model with $H/R=0.05$. In addition, a disk in the diffusive regime is also subject to instability that may enhance its breaking \citep{Dogan2018}. This effect may also help the $H/R=0.02$ model break, although it is not well within the viscous regime.

In this letter, we show that a circumtriple disk with an aspect ratio of $H/r = 0.02$ breaks due to the influence of  the triple stars with the properties of GW Ori, whereas a disk with $H/r = 0.05$ does not break. However, for observationally motivated parameters of protoplanetary disks, a value of $H/r = 0.02$ is unlikely. In Section~\ref{sec::setup}, we detail the setup of the hydrodynamical simulations and present the results in Section~\ref{sec::results}. In Section~\ref{sec::discussion}, we discuss possible aspect ratios for the protoplanetary disk around GW Ori. Finally, in Section~\ref{sec::conclusion}, we share our conclusions.

 \begin{table}
 	\caption{Summary of the hydrodynamical simulations. The first column denotes the model name. The second column denotes the initial disk tilt with respect to the triple star orbital plane, $i_0$. The third column denotes the initial disk aspect ratio at the inner disk edge, $H/r$. The last column denotes whether disk breaking occurs.}
  \begin{tabular}{lccc} 
 		\hline
    Model & $i_0$ & $H/r$ & disk breaking\\
 	    \hline
         \hline
         i38\_h0p02 & $38^\circ$ & $0.02$ & yes\\
         i38\_h0p05 & $38^\circ$ & $0.05$ & no\\
         i28\_h0p02 & $28^\circ$ & $0.02$ & yes\\
         i28\_h0p05 & $28^\circ$ & $0.05$ & no\\
	\hline
 \label{table::sims}
 \end{tabular}
  \end{table}


\section{Simulation setup}
\label{sec::setup}
For our numerical simulations, we use the 3D smoothed particle hydrodynamics (SPH) code {\sc Phantom} \citep{Price2012,Price2010,Lodato2010,Price2018}. We simulate a misaligned accretion disk around the triple star system GW Ori. The {\sc Phantom} code has been extensively tested and applied to model misaligned accretion disks in triple systems \citep[e.g.,][]{Smallwood2021,Ceppi2022}.

We adopt the triple star system parameters from \citet{Kraus2020}.  The mass of the inner binary A-B is $M_{\rm A} = 2.47\pm 0.33\, \rm M_{\odot}$ and $M_{\rm B} = 1.43 \pm 0.18\, \rm M_{\odot}$, while the mass of the tertiary companion C is $M_{\rm C} = 1.36\pm 0.28\, \rm M_{\odot}$. The separation of the inner binary is $a_{\rm A-B} = 1.20 \pm 0.04 \, \rm au$, while the separation of the tertiary is $a_{\rm AB-C} = 8.89 \pm 0.04 \, \rm au$. The inner binary and tertiary star are initialized at apastron.  Since we are not modeling material close to the A-B binary, we increase the accretion radii of stars A and B to $r_{\rm acc} = 0.5\, \rm au$ ($\sim 0.42a_{\rm A-B}$) to enhance computational efficiency. Similarly, the accretion radius of the tertiary companion is set to $r_{\rm acc} = 2.3\, \rm au$, approximately $0.25a_{\rm AB-C}$, to also improve computational efficiency.
Particles entering these accretion radii are considered accreted, with their mass and angular momentum added to the respective star \citep[e.g.][]{Bate1995}.

Each simulation employs $5\times10^5$ equal-mass Lagrangian particles, distributed between the inner disk radius, $r_{\rm in}$, and the outer disk radius, $r_{\rm out}$. The observed outer radius of the gas disk is approximately $1300\, \rm au$ \citep{Bi2020}, suggesting that most of the angular momentum is concentrated in the outer regions. To improve computational efficiency and resolution, we truncate the outer radius to $r_{\rm out}=200\, \rm au$, similar to \cite{Kraus2020}. This truncation still maintains the angular momentum distribution, with the outer disk holding the majority of the angular momentum.  We run the simulations for $2000\, \rm P_{orb}$, where $P_{\rm orb}$ represents the orbital period of the outer binary, equivalent to approximately $2.3\times 10^{4}\, \rm yr$. In comparison, the disk structure in \cite{Kraus2020} is shown only at $9500\, \rm yr$.

We model two initial disk misalignments with respect to the triple star orbital plane: $38^\circ$ and $28^\circ$. The tilt of $38^\circ$ was first reported in \cite{Kraus2020}. However, reanalysis of the system's orientation revealed that the disk is tilted by $28^\circ$ relative to the triple star orbital plane. This updated tilt value is discussed in Section 4.4 of \cite{Young2023}. For comparison, we model both misalignments. A summary of the simulations are given in Table~\ref{table::sims}.

The total mass of the disk is set to $0.1\, \rm M_{\odot}$, based on a dust-to-gas ratio of $0.01$ and observations of the dust mass \cite[e.g.,][]{Bi2020}. \cite{Kraus2020} estimated a lower disk mass due to incomplete flux recovery from their ALMA observations, which lacked short baselines. We ignore the effects of self-gravity, as it does not affect the nodal precession rate of flat circumbinary disks and the disk mass is not large enough for self-gravity to be a significant factor.

\begin{figure*} 
\includegraphics[width=0.52\columnwidth]{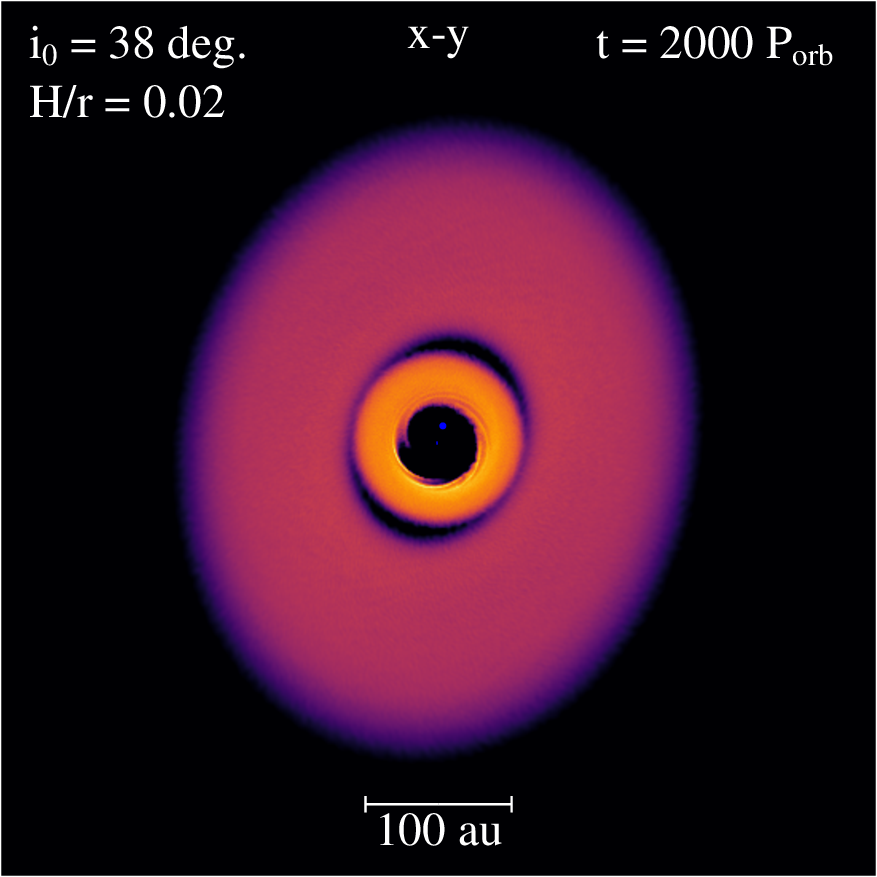}
\includegraphics[width=0.52\columnwidth]{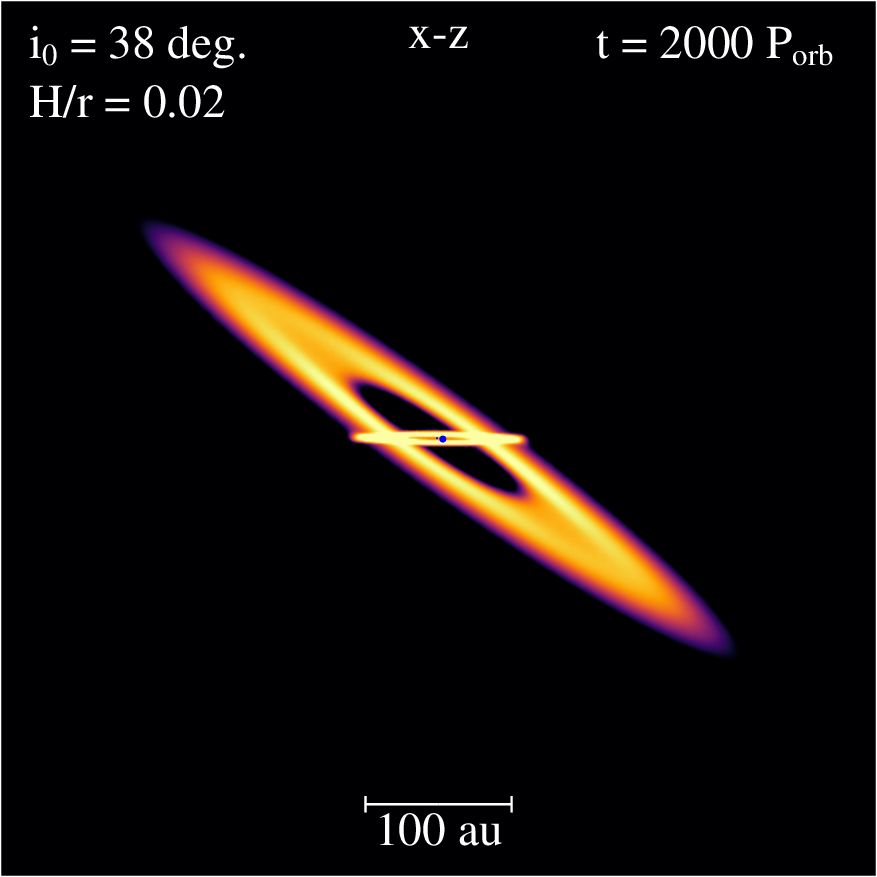}
\includegraphics[width=0.52\columnwidth]{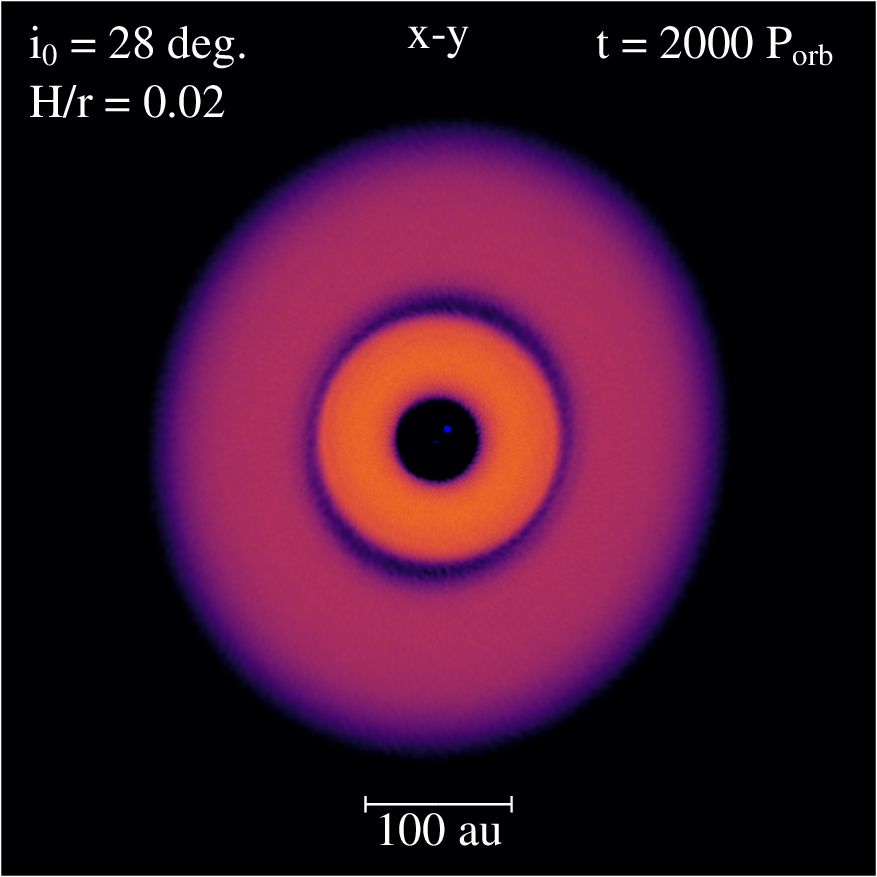}
\includegraphics[width=0.52\columnwidth]{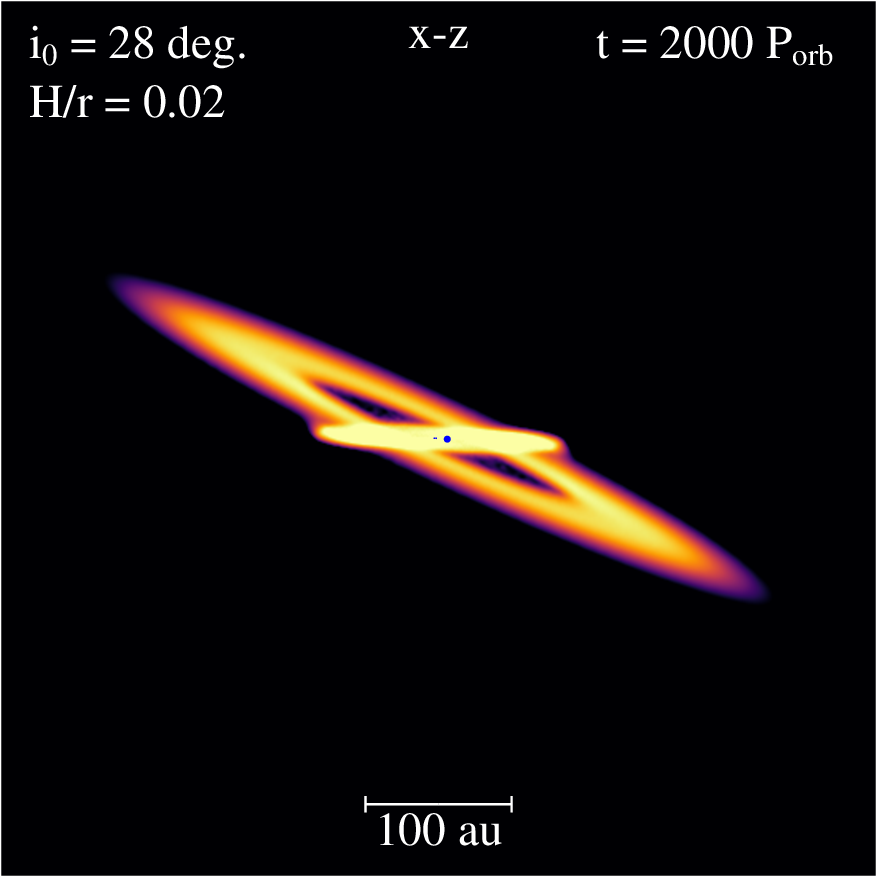}
\includegraphics[width=0.52\columnwidth]{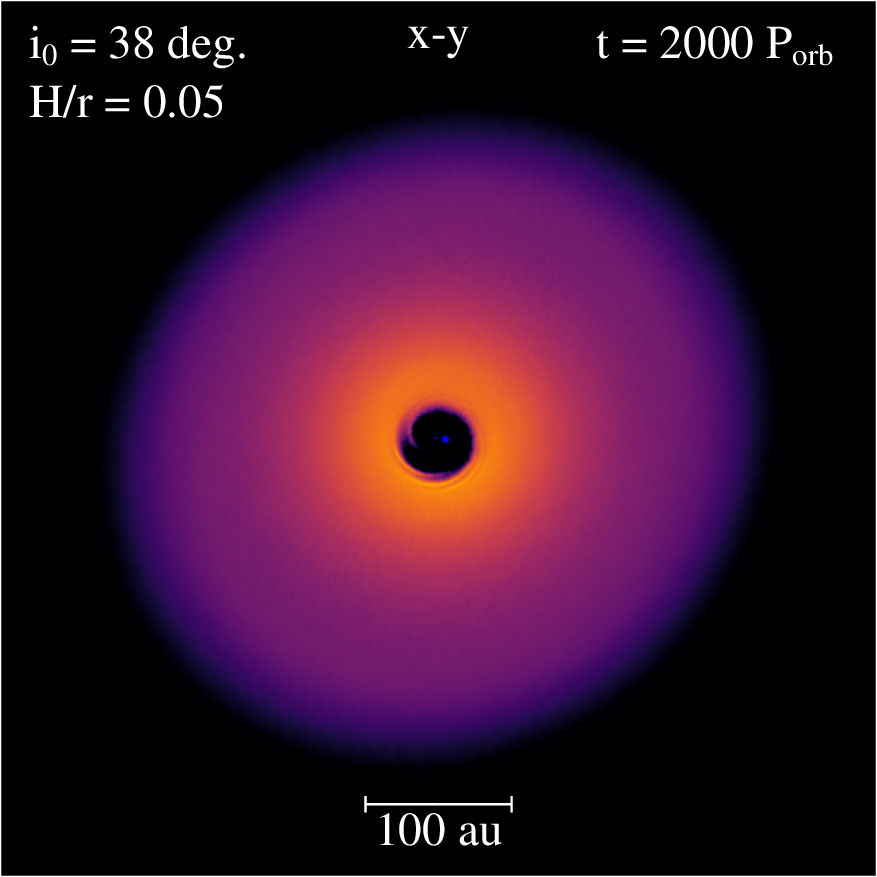}
\includegraphics[width=0.52\columnwidth]{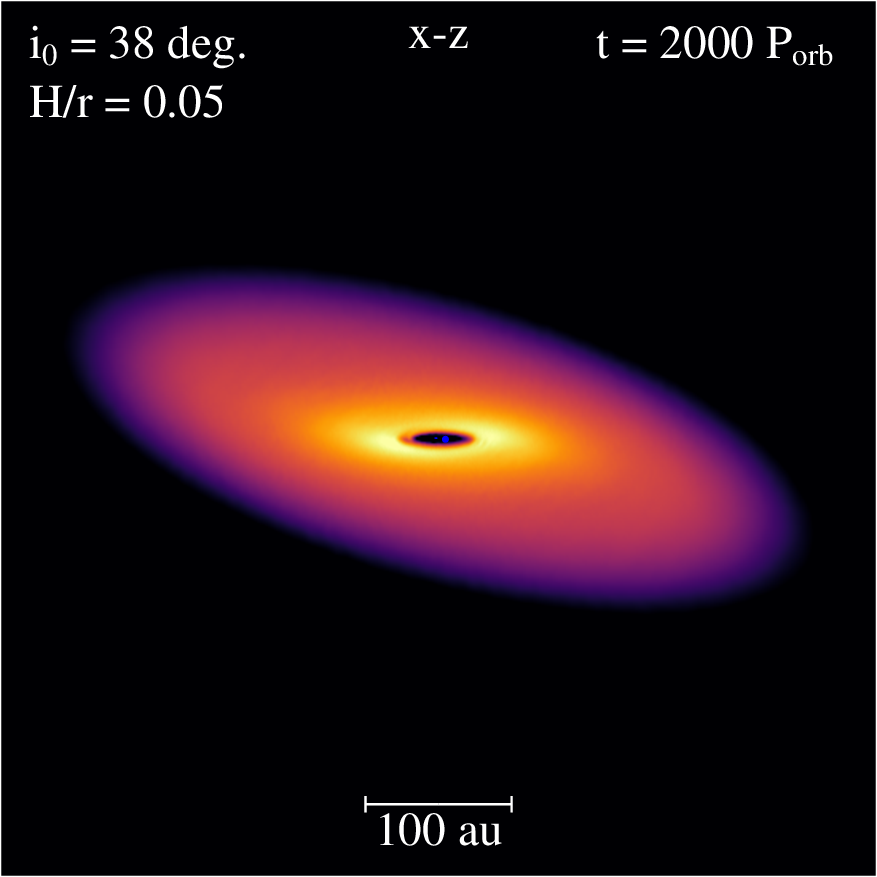}
\includegraphics[width=0.52\columnwidth]{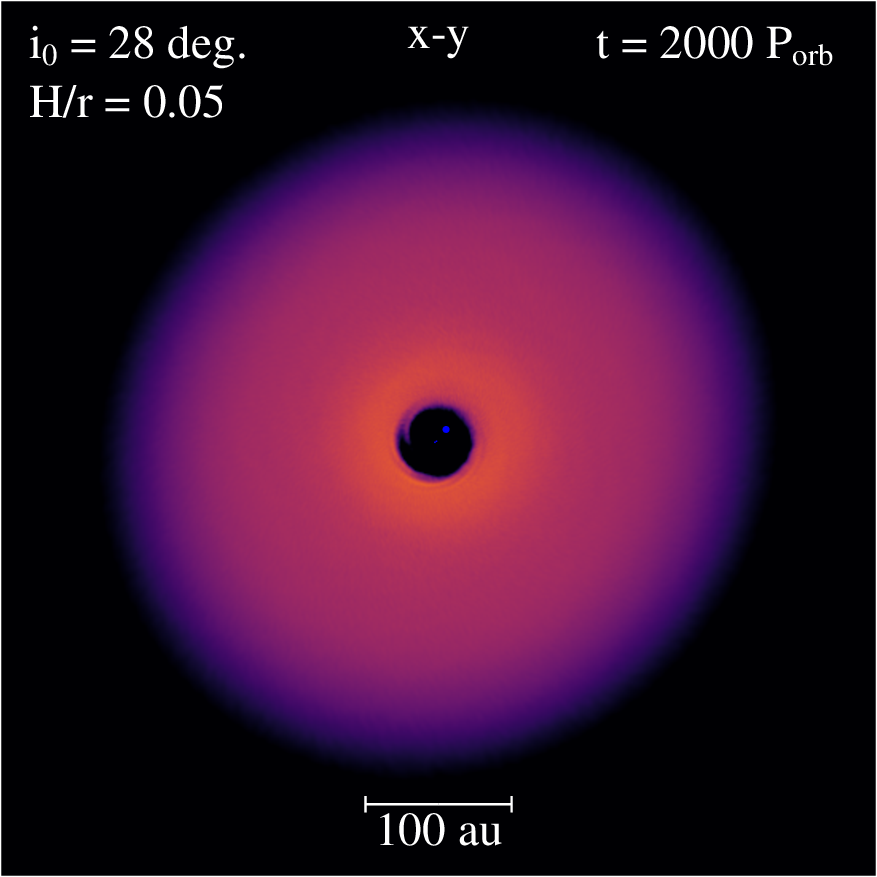}
\includegraphics[width=0.52\columnwidth]{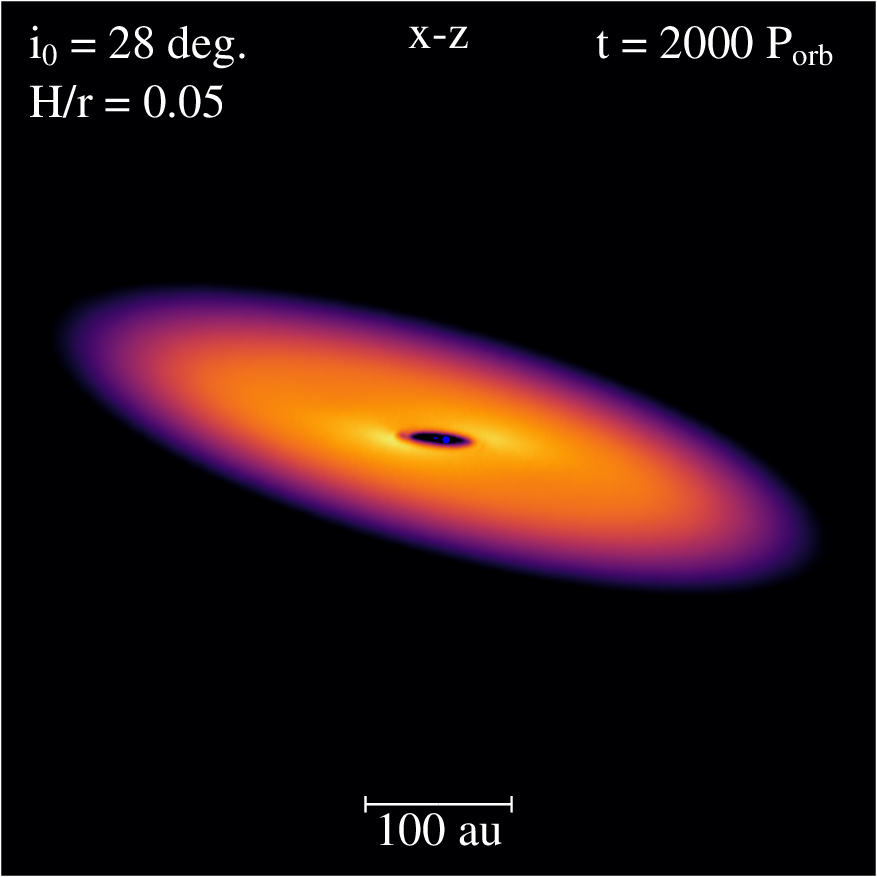}
\centering
\caption{Circumtriple disk evolution around GW Ori with an initial misalignment between the disk and the outer binary of the triple star of $i_0 = 38^\circ$ (first two columns) and $i_0 = 28^\circ$ (last two columns) at time $t = 2000\, \rm P_{orb}$. The upper row represents disks with $H/r = 0.02$, while the bottom row shows those with $H/r = 0.05$. The triple stars are depicted by blue dots. Columns 1 and 3 show the view looking down onto the outer binary orbital plane (the $x$--$y$ plane), while columns 2 and 4 present the $x$–$z$ plane, with the AB-C binary eccentricity aligned along the $x$--axis. The color indicates gas density, with yellow regions being about two orders of magnitude higher than the purple regions.}
\label{fig::splash_i38}
\end{figure*}



The initial surface density profile of the disk follows a power law distribution given by:
\begin{equation}
\Sigma(r) = \Sigma_0 \bigg( \frac{r}{r_{\rm in}} \bigg)^{-p},
\label{eq::sigma}
\end{equation}
where $\Sigma_0$ is the density normalization, $r$ is the spherical disc radius, and $p$ is the power law index. We set $p=0.5$. The normalization $\Sigma_0$ is determined based on the total disk mass, which approximates the mass of the three dust rings inferred by \cite{Bi2020}, assuming a gas-to-dust ratio of 100.
We model the disk as locally isothermal (the temperature is measured from the center-of-mass), with a hydrostatic thickness that scales with radius as:
\begin{equation}
H = \frac{c_{\rm s}}{\Omega} \propto r^{3/2-q},
\end{equation}
where $\Omega = \sqrt{GM/r^3}$ is the orbital frequency and $c_{\rm s}$ is the sound speed which are independent of time. To maintain a constant aspect ratio $H/r$ across the disk, we set $q = 0.5$.

The $\alpha$ viscosity parameter is calculated using the prescription from \cite{Lodato2010}:
\begin{equation}
\alpha \approx \frac{\alpha_{\rm AV}}{10} \frac{\langle h \rangle}{H},
\end{equation}
where $\alpha_{\rm AV}$ represents the artificial viscosity and $\langle h \rangle$ is the mean smoothing length of particles in a cylindrical ring at a given radius \citep{Lodato2010}. The simulations from \cite{Kraus2020} used a viscosity ranging from $\approx 0.01 - 0.02$. Therefore, we set $\alpha \approx 0.015$\footnote{Observations have shown that $\alpha$ in protoplanetary disks can be as low as $10^{-3}$ to $10^{-4}$ \cite[e.g.,][]{Flaherty2015,Teague2016,Pinte2016,Flaherty2018,Lesur2023}. However, adopting this higher $\alpha$ value does not affect the conclusions, as the disk is already in the bending wave regime under our initial conditions.}. We examine two values for the disk aspect ratio at $r = r_{\rm in}$: $H/r = 0.02$ and $0.05$.  Compared to $\alpha \approx 0.015$, a disk with $H/r = 0.02$ (as noted in \cite{Kraus2020}) lies near the boundary of the diffusive regime, whereas a disk with $H/r = 0.05$ falls within the wave-like regime. Since the disk aspect ratio is constant, both viscosity and mean smoothing length vary across the disk. The shelled-averaged smoothing length per scale height for the simulations is initially $\langle h \rangle/H \approx 0.70$,  this is well enough resolved to capture breaking \cite[e.g.,][]{Nealon2015}.
\begin{figure*} 
\centering
\includegraphics[width=\columnwidth]{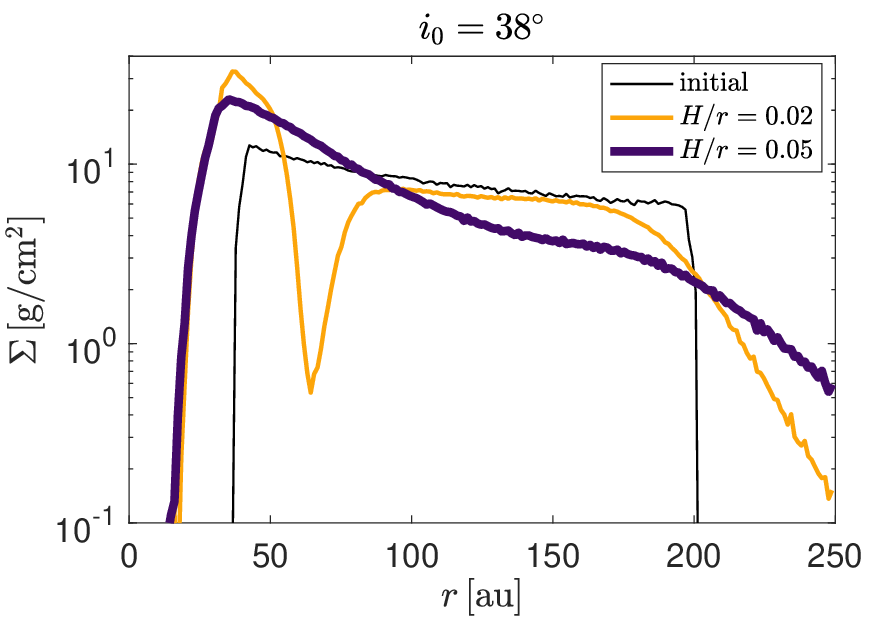}
\includegraphics[width=\columnwidth]{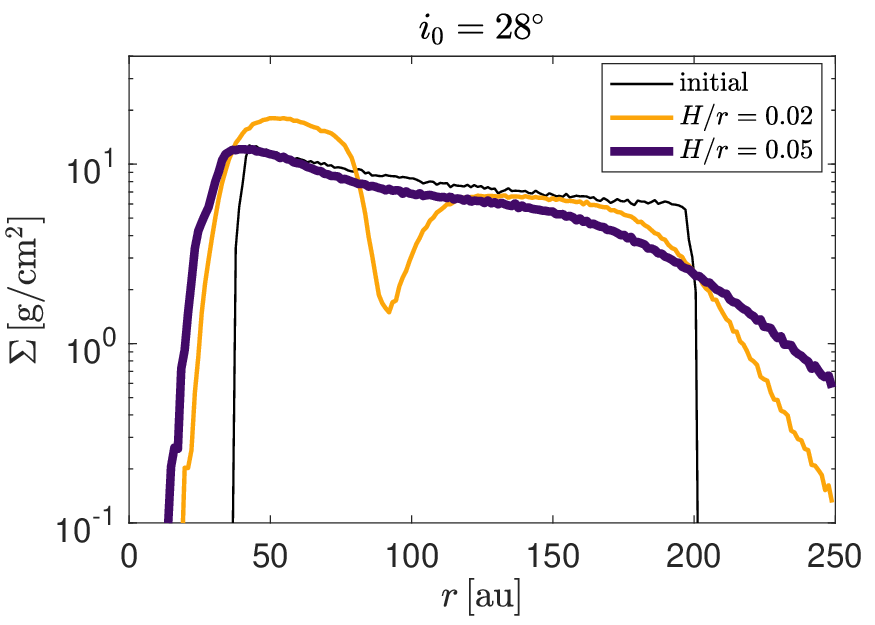}
\caption{The disk surface density, $\Sigma$ as a function of radius, $r$, for initial circumtriple disk misalignment of $i_0 = 38^\circ$ (left plot) and $i_0 = 28^\circ$ (right plot). The initial surface density profile is given by the flat black line. The thin yellow and thick purple  lines denote $H/r = 0.02$ and $H/r = 0.05$, respectively. For each disk misalignment, the disk breaks when $H/r = 0.02$. }
\label{fig::sigma}
\end{figure*}


\section{Results}
\label{sec::results}

Figure~\ref{fig::splash_i38} shows the three-dimensional structure of the circumtriple disk around GW Ori for initial disk misalignment to the outer binary orbit of the triple star of $i_0 = 38^\circ$ (first two columns) and $i_0 = 28^\circ$ (last two columns) at time $t = 2000\, \rm P_{orb}$, where $P_{\rm orb}$ is the orbital period of the outer binary orbit. The upper row represents disks with $H/r = 0.02$, while the bottom row shows those with $H/r = 0.05$. For $H/r = 0.02$ with $i_0 = 38^\circ$, the disk breaks due to the torque from the triple stars, while for the thicker disk ($H/r = 0.05$), it does not break. Similarly, for an initial disk misalignment of $i_0 = 28^\circ$, the thinner disk ($H/r = 0.02$) breaks, while the thicker disk ($H/r = 0.05$) shows no evidence of breaking. 

 Fig.~\ref{fig::sigma} shows the surface density profile as a function of disk radius for $i_0 = 38^\circ$ (right) and $i_0 = 28^\circ$ (left) at a time of $2000\, \rm P_{orb}$. The initial power-law index is set to $p = 1.5$, shown by the black curve. The yellow and purple curves denote the surface density profiles for $H/r = 0.02$ and $H/r = 0.05$, respectively. The dip in the surface density profile for $H/r = 0.02$ corresponds to the location of the break. At the end of the simulation, for the initially higher-inclined disk, $i_0 = 38^\circ$, the break is located at $r \sim 70\, \rm au$. For the lower initial tilt, $i_0 = 28^\circ$, the break is located at $r \sim 95\, \rm au$. The breaking radius for the $i_0 = 28^\circ$ disk is farther out than that of the $i_0 = 38^\circ$ disk at the same timestep. Furthermore, the gap in the $i_0 = 38^\circ$ disk is deeper than the gap from the more lowly inclined disk. From the surface density profiles for a thicker disk, $H/r = 0.05$, there is no evidence of disk breaking, meaning the surface density profile is relatively smooth. A summary of when disk breaking occurs from the simulations are given in the last column in Table~\ref{table::sims}. 



\begin{figure} 
\includegraphics[width=\columnwidth]{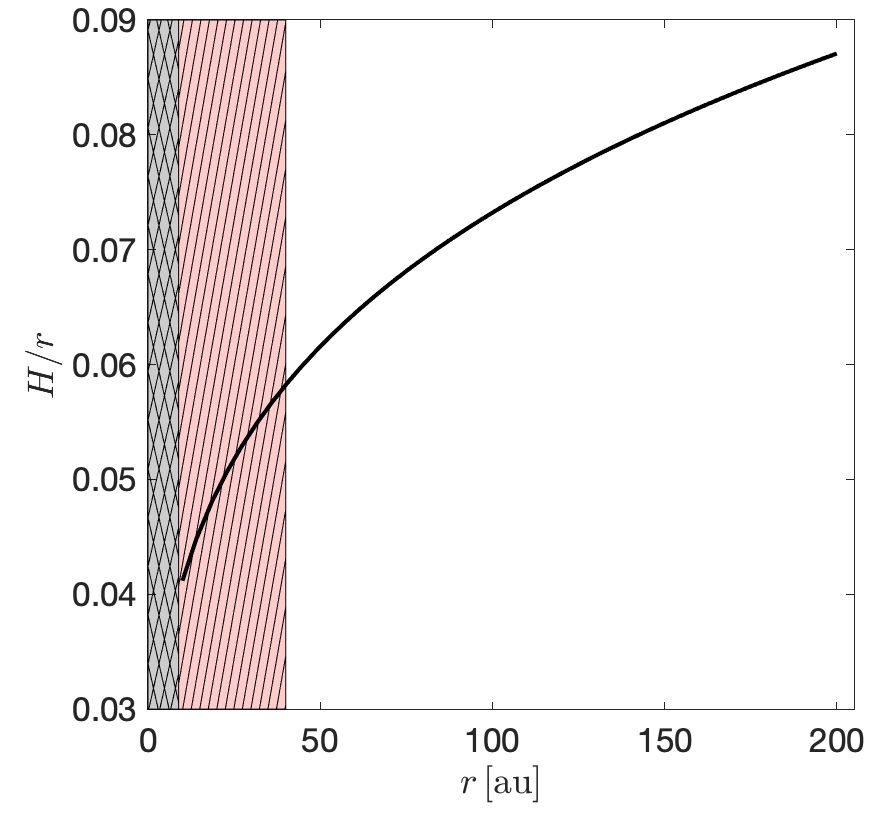}
\centering
\caption{The observationally estimated disk aspect ratio, $H/r$, as a function of disk radius, $r$, for GW Ori from Equation~\ref{eq::scaleheight}. The black double-hatched area represents the orbital extent of the AB-C binary. The red single-hatched area denotes the area where the tidal torque truncates the disk.  }
\label{fig::H_r}
\end{figure}

\section{discussion}
\label{sec::discussion}
The hydrodynamical simulations show that a disk with an initial aspect ratio of $H/r = 0.02$ breaks due to interactions with the GW Ori triple star system. However, a thicker disk with $H/r = 0.05$ shows no signs of breaking. As mentioned in the introduction, the original version of \cite{kraus2020v1} used an aspect ratio of $H/r = 0.05$ and reported disk breaking in their simulations. However, in the revised version of \cite{Kraus2020} (version 2), the aspect ratio was changed to $H/r = 0.02$ without altering any of the hydrodynamical results from version 1. Here, we argue, based on observational evidence, that disk aspect ratios as thin as $H/r = 0.02$ are unlikely for protoplanetary disks.

A temperature as a function of radius for a simple irradiated flaring disk is given by
\begin{equation}
    T_{\rm d}(r) = \bigg(\frac{\frac{1}{2}\varphi L_{\star}}{4\pi r^2 \sigma_{\rm SB}} \bigg)^{1/4},
\end{equation}
where $L_{\star}$ is the bolometric luminosity, $\sigma_{\rm SB}$ is the Stefan-Boltzmann constant, and $\varphi$ is the flaring angle \cite[e.g.,][]{Chiang1997,DAlessio1998,Dullemond2001,Dullemond2018}. Based on previous models, we set the flaring angle to be $\varphi = 0.02$. The bolometric luminosity for GW Ori is $L_{\star} \sim 48\, \rm L_{\odot}$, assuming that the primary dominates the total emission of the system \citep{Fang2014,Bi2020,Kraus2020}. GW Ori C is much less luminous than GW Ori A \citep{Berger2011}. The pressure scale height of the disk is given by
\begin{equation}
    H = \sqrt{\frac{k_{\rm B} T_{\rm d} r^3}{\mu m_p G M_{\star}}}, 
    \label{eq::scaleheight}
\end{equation}
where $k_{\rm B}$ is the Boltzmann constant, $m_p$ is the proton mass, $G$ is the gravitational constant, and $\mu = 2.3$ being the mean molecular weight in atomic units.

Figure~\ref{fig::H_r} shows the disk aspect ratio from Equation~\ref{eq::scaleheight} as a function of disk radius for GW Ori. The black double-hatched area represents the orbital extent of the AB-C binary, and the red single-hatched area denotes where the tidal torque truncates the disk \cite[e.g.,][]{Artymowicz1994}. At $40\, \rm au$ (the initial inner disk edge from the SPH simulations), the disk aspect ratio is $H/r \sim 0.06$.  An  aspect ratio of $H/r = 0.05$ at $r = 40\, \rm au$ does not lead to disk breaking under the torque of the triple star system.  Even if the inner edge of the disk is less than $40\, \rm au$, $H/r$ does not reach a value of $0.02$, which is the value used in \cite{Kraus2020}.  Given that we assume a constant $H/r$ in the simulations, the actual disk of GW Ori is likely to be thicker than our simulations suggest, making it more resistant to disk breaking. The typical values of the disk aspect ratio for protoplanetary disks are $H/r > 0.05$ \cite[e.g.,][]{Hartmann1998}.

 \cite{Smallwood2021} proposed an alternative explanation for the prominent disk gap observed in GW Ori, attributing it to the presence of one or more planets within the system. Their findings suggest that a massive planet, initially misaligned with the disk, could repeatedly open a gap within a thin disk ($H/r \approx 0.05$) as its orbital inclination oscillates in and out of the disk plane. Over time, the broken inner disk precesses and aligns with the orbital plane of the stars, while the outer disk remains misaligned—consistent with the observed disk structure. 
Conversely, earlier studies \citep[e.g.,][]{Dipierro2016} indicate that a low-mass planet, tightly coupled to the gas, could still produce a gap in the dust component of the disk. 
Furthermore, synthetic CO moment 1 maps generated by \cite{Smallwood2021} show that a model incorporating a circumtriple planet provides a better match to the observations by \cite{Bi2020} than a planet-free model. These results collectively challenge the hypothesis that the break in GW Ori’s circumtriple disk arises from stellar torques acting on the disk. Instead, they point to the presence of undetected planets as the likely cause of the observed disk-breaking phenomenon, potentially marking the first detection of planets in a circumtriple orbit.

 Disk misalignment relative to the stellar orbital plane may arise from the infall of material in the form of gaseous streamers with varying angular momentum, which may contribute to disk warping. For example, the observed disk misalignment in SU Aur may be linked to the presence of infalling material \citep{Ginski2021}. Additional examples of observed misaligned disks associated with infall processes have been reported  by \cite{Garufi2024}. Beyond observational data, theoretical studies bolster this explanation, demonstrating that infalling material can induce primordial misalignments during the early stages of disk formation \citep{Thies2011,Bate2018,Kuffmeier2021}. Moreover, larger-scale simulations indicate that infall from the surrounding environment is frequently misaligned relative to the orientation of pre-existing disks, emphasizing the prevalence of such dynamics in star-forming regions \citep{Kuffmeier2024,Pelkonen2024}. The primordial misalignment of the circumtriple disk around GW Ori was likely caused by the infall of material with angular momenta misaligned relative to the plane of the triple stars.  Multiple episodes of disk formation is
another possible explanation for the disk
orientations in GW Ori.


\section{Conclusion}
\label{sec::conclusion}
In this letter, we used hydrodynamical simulations to model the circumtriple disk around GW Ori to further investigate the origin of the observed gap. Previous models by \cite{Kraus2020} suggested that the gap is produced by the interaction of the disk with the triple stars, causing tearing and eventually breaking the disk. However, they modeled an extremely thin disk with an aspect ratio of $H/r \sim 0.02$, which places their model near the diffusive regime. Protoplanetary disks are more likely to have thicker disks, with $H/r \gtrsim 0.05$, and to be in the bending wave regime \cite[e.g.,][]{Hartmann1998}. The simulations conducted in this letter show that with a circumtriple disk having $H/r = 0.05$, the disk does not break. A simple calculation of the pressure scale height for the circumtriple disk around GW Ori suggests that the aspect ratio should be larger than $0.05$. Therefore, an alternative scenario for the gap's origin may be the presence of circumtriple planets orbiting GW Ori \citep{Smallwood2021}.

\begin{acknowledgments}
 We thank the anonymous referee for useful comments.  The authors also thank Alison Young for comments.  JLS acknowledges funding from the ASIAA Distinguished Postdoctoral Fellowship, the Taiwan Foundation for the Advancement of Outstanding Scholarship and the Dodge Family Prize Fellowship in Astrophysics. We acknowledge support from NASA through grants 80NSSC21K0395 and 80NSSC19K0443. RN acknowledges funding from UKRI/EPSRC through a Stephen Hawking Fellowship (EP/T017287/1).
\end{acknowledgments}


 \bibliography{ref}{}

\begin{thebibliography}{}
\expandafter\ifx\csname natexlab\endcsname\relax\def\natexlab#1{#1}\fi
\providecommand{\url}[1]{\href{#1}{#1}}
\providecommand{\dodoi}[1]{doi:~\href{http://doi.org/#1}{\nolinkurl{#1}}}
\providecommand{\doeprint}[1]{\href{http://ascl.net/#1}{\nolinkurl{http://ascl.net/#1}}}
\providecommand{\doarXiv}[1]{\href{https://arxiv.org/abs/#1}{\nolinkurl{https://arxiv.org/abs/#1}}}

\bibitem[{{Artymowicz} \& {Lubow}(1994)}]{Artymowicz1994}
{Artymowicz}, P., \& {Lubow}, S.~H. 1994, \apj, 421, 651,
  \dodoi{10.1086/173679}

\bibitem[{{Bate}(2018)}]{Bate2018}
{Bate}, M.~R. 2018, \mnras, 475, 5618, \dodoi{10.1093/mnras/sty169}

\bibitem[{{Bate} {et~al.}(1995){Bate}, {Bonnell}, \& {Price}}]{Bate1995}
{Bate}, M.~R., {Bonnell}, I.~A., \& {Price}, N.~M. 1995, \mnras, 277, 362,
  \dodoi{10.1093/mnras/277.2.362}

\bibitem[{{Benz} {et~al.}(1990){Benz}, {Bowers}, {Cameron}, \&
  {Press}}]{Benz1990}
{Benz}, W., {Bowers}, R.~L., {Cameron}, A.~G.~W., \& {Press}, W.~H.~. 1990,
  \apj, 348, 647, \dodoi{10.1086/168273}

\bibitem[{{Berger} {et~al.}(2011){Berger}, {Monnier}, {Millan-Gabet}, {Renard},
  {Pedretti}, {Traub}, {Bechet}, {Benisty}, {Carleton}, {Haguenauer}, {Kern},
  {Labeye}, {Longa}, {Lacasse}, {Malbet}, {Perraut}, {Ragland}, {Schloerb},
  {Schuller}, \& {Thi{\'e}baut}}]{Berger2011}
{Berger}, J.~P., {Monnier}, J.~D., {Millan-Gabet}, R., {et~al.} 2011, \aap,
  529, L1, \dodoi{10.1051/0004-6361/201016219}

\bibitem[{{Bi} {et~al.}(2020){Bi}, {van der Marel}, {Dong}, {Muto}, {Martin},
  {Smallwood}, {Hashimoto}, {Liu}, {Nomura}, {Hasegawa}, {Takami}, {Konishi},
  {Momose}, {Kanagawa}, {Kataoka}, {Ono}, {Sitko}, {Takahashi}, {Tomida}, \&
  {Tsukagoshi}}]{Bi2020}
{Bi}, J., {van der Marel}, N., {Dong}, R., {et~al.} 2020, \apjl, 895, L18,
  \dodoi{10.3847/2041-8213/ab8eb4}

\bibitem[{{Ceppi} {et~al.}(2022){Ceppi}, {Cuello}, {Lodato}, {Clarke}, {Toci},
  \& {Price}}]{Ceppi2022}
{Ceppi}, S., {Cuello}, N., {Lodato}, G., {et~al.} 2022, \mnras, 514, 906,
  \dodoi{10.1093/mnras/stac1390}

\bibitem[{{Chiang} \& {Goldreich}(1997)}]{Chiang1997}
{Chiang}, E.~I., \& {Goldreich}, P. 1997, \apj, 490, 368,
  \dodoi{10.1086/304869}

\bibitem[{{D'Alessio} {et~al.}(1998){D'Alessio}, {Cant{\"o}}, {Calvet}, \&
  {Lizano}}]{DAlessio1998}
{D'Alessio}, P., {Cant{\"o}}, J., {Calvet}, N., \& {Lizano}, S. 1998, \apj,
  500, 411, \dodoi{10.1086/305702}

\bibitem[{{Dipierro} {et~al.}(2016){Dipierro}, {Laibe}, {Price}, \&
  {Lodato}}]{Dipierro2016}
{Dipierro}, G., {Laibe}, G., {Price}, D.~J., \& {Lodato}, G. 2016, \mnras, 459,
  L1, \dodoi{10.1093/mnrasl/slw032}

\bibitem[{{Dong} {et~al.}(2018){Dong}, {Li}, {Chiang}, \& {Li}}]{Dong2018}
{Dong}, R., {Li}, S., {Chiang}, E., \& {Li}, H. 2018, \apj, 866, 110,
  \dodoi{10.3847/1538-4357/aadadd}

\bibitem[{{Do{\v{g}}an} {et~al.}(2018){Do{\v{g}}an}, {Nixon}, {King}, \&
  {Pringle}}]{Dogan2018}
{Do{\v{g}}an}, S., {Nixon}, C.~J., {King}, A.~R., \& {Pringle}, J.~E. 2018,
  \mnras, 476, 1519, \dodoi{10.1093/mnras/sty155}

\bibitem[{{Dullemond} {et~al.}(2001){Dullemond}, {Dominik}, \&
  {Natta}}]{Dullemond2001}
{Dullemond}, C.~P., {Dominik}, C., \& {Natta}, A. 2001, \apj, 560, 957,
  \dodoi{10.1086/323057}

\bibitem[{{Dullemond} {et~al.}(2018){Dullemond}, {Birnstiel}, {Huang},
  {Kurtovic}, {Andrews}, {Guzm{\'a}n}, {P{\'e}rez}, {Isella}, {Zhu}, {Benisty},
  {Wilner}, {Bai}, {Carpenter}, {Zhang}, \& {Ricci}}]{Dullemond2018}
{Dullemond}, C.~P., {Birnstiel}, T., {Huang}, J., {et~al.} 2018, \apjl, 869,
  L46, \dodoi{10.3847/2041-8213/aaf742}

\bibitem[{{Facchini} {et~al.}(2013){Facchini}, {Lodato}, \&
  {Price}}]{Facchini2013}
{Facchini}, S., {Lodato}, G., \& {Price}, D.~J. 2013, \mnras, 433, 2142,
  \dodoi{10.1093/mnras/stt877}

\bibitem[{{Fang} {et~al.}(2014){Fang}, {Sicilia-Aguilar}, {Roccatagliata},
  {Fedele}, {Henning}, {Eiroa}, \& {M{\"u}ller}}]{Fang2014}
{Fang}, M., {Sicilia-Aguilar}, A., {Roccatagliata}, V., {et~al.} 2014, \aap,
  570, A118, \dodoi{10.1051/0004-6361/201424146}

\bibitem[{{Flaherty} {et~al.}(2015){Flaherty}, {Hughes}, {Rosenfeld},
  {Andrews}, {Chiang}, {Simon}, {Kerzner}, \& {Wilner}}]{Flaherty2015}
{Flaherty}, K.~M., {Hughes}, A.~M., {Rosenfeld}, K.~A., {et~al.} 2015, \apj,
  813, 99, \dodoi{10.1088/0004-637X/813/2/99}

\bibitem[{{Flaherty} {et~al.}(2018){Flaherty}, {Hughes}, {Teague}, {Simon},
  {Andrews}, \& {Wilner}}]{Flaherty2018}
{Flaherty}, K.~M., {Hughes}, A.~M., {Teague}, R., {et~al.} 2018, \apj, 856,
  117, \dodoi{10.3847/1538-4357/aab615}

\bibitem[{{Garufi} {et~al.}(2024){Garufi}, {Ginski}, {van Holstein}, {Benisty},
  {Manara}, {P{\'e}rez}, {Pinilla}, {Ribas}, {Weber}, {Williams}, {Cieza},
  {Dominik}, {Facchini}, {Huang}, {Zurlo}, {Bae}, {Hagelberg}, {Henning},
  {Hogerheijde}, {Janson}, {M{\'e}nard}, {Messina}, {Meyer}, {Pinte}, {Quanz},
  {Rigliaco}, {Roccatagliata}, {Schmid}, {Szul{\'a}gyi}, {van Boekel},
  {Wahhaj}, {Antichi}, {Baruffolo}, \& {Moulin}}]{Garufi2024}
{Garufi}, A., {Ginski}, C., {van Holstein}, R.~G., {et~al.} 2024, \aap, 685,
  A53, \dodoi{10.1051/0004-6361/202347586}

\bibitem[{{Ginski} {et~al.}(2021){Ginski}, {Facchini}, {Huang}, {Benisty},
  {Vaendel}, {Stapper}, {Dominik}, {Bae}, {M{\'e}nard}, {Muro-Arena},
  {Hogerheijde}, {McClure}, {van Holstein}, {Birnstiel}, {Boehler}, {Bohn},
  {Flock}, {Mamajek}, {Manara}, {Pinilla}, {Pinte}, \& {Ribas}}]{Ginski2021}
{Ginski}, C., {Facchini}, S., {Huang}, J., {et~al.} 2021, \apjl, 908, L25,
  \dodoi{10.3847/2041-8213/abdf57}

\bibitem[{{Hartmann} {et~al.}(1998){Hartmann}, {Calvet}, {Gullbring}, \&
  {D'Alessio}}]{Hartmann1998}
{Hartmann}, L., {Calvet}, N., {Gullbring}, E., \& {D'Alessio}, P. 1998, \apj,
  495, 385, \dodoi{10.1086/305277}

\bibitem[{Kraus(2020)}]{kraus2020media}
Kraus, S. 2020, arXiv preprint arXiv:2012.06578

\bibitem[{{Kraus} {et~al.}(2020){Kraus}, {Kreplin}, {Young}, {Bate}, {Monnier},
  {Harries}, {Avenhaus}, {Kluska}, {Laws}, {Rich}, {Willson}, {Aarnio},
  {Adams}, {Andrews}, {Anugu}, {Bae}, {ten Brummelaar}, {Calvet}, {Cur{\'e}},
  {Davies}, {Ennis}, {Espaillat}, {Gardner}, {Hartmann}, {Hinkley}, {Labdon},
  {Lanthermann}, {LeBouquin}, {Schaefer}, {Setterholm}, {Wilner}, \&
  {Zhu}}]{Kraus2020}
{Kraus}, S., {Kreplin}, A., {Young}, A.~K., {et~al.} 2020, arXiv e-prints,
  arXiv:2004.01204.
\newblock \doarXiv{2004.01204}

\bibitem[{Kraus {et~al.}(2020 v1)Kraus, Kreplin, Young, Bate, Monnier, Harries,
  Avenhaus, Kluska, Laws, Rich, Willson, Aarnio, Adams, Andrews, Anugu, Bae,
  ten Brummelaar, Calvet, Curé, Davies, Ennis, Espaillat, Gardner, Hartmann,
  Hinkley, Labdon, Lanthermann, LeBouquin, Schaefer, Setterholm, Wilner, \&
  Zhu}]{kraus2020v1}
Kraus, S., Kreplin, A., Young, A.~K., {et~al.} 2020 v1, A triple star system
  with a misaligned and warped circumstellar disk shaped by disk tearing,
  \dodoi{https://doi.org/10.1126/science.aba4633}

\bibitem[{{Kuffmeier} {et~al.}(2021){Kuffmeier}, {Dullemond}, {Reissl}, \&
  {Goicovic}}]{Kuffmeier2021}
{Kuffmeier}, M., {Dullemond}, C.~P., {Reissl}, S., \& {Goicovic}, F.~G. 2021,
  \aap, 656, A161, \dodoi{10.1051/0004-6361/202039614}

\bibitem[{{Kuffmeier} {et~al.}(2024){Kuffmeier}, {Pineda}, {Segura-Cox}, \&
  {Haugb{\o}lle}}]{Kuffmeier2024}
{Kuffmeier}, M., {Pineda}, J.~E., {Segura-Cox}, D., \& {Haugb{\o}lle}, T. 2024,
  \aap, 690, A297, \dodoi{10.1051/0004-6361/202450410}

\bibitem[{{Larwood} \& {Papaloizou}(1997)}]{Larwood1997}
{Larwood}, J.~D., \& {Papaloizou}, J.~C.~B. 1997, \mnras, 285, 288,
  \dodoi{10.1093/mnras/285.2.288}

\bibitem[{{Lepp} {et~al.}(2023){Lepp}, {Martin}, \& {Lubow}}]{Lepp2023}
{Lepp}, S., {Martin}, R.~G., \& {Lubow}, S.~H. 2023, \apjl, 943, L4,
  \dodoi{10.3847/2041-8213/acaf6d}

\bibitem[{{Lesur} {et~al.}(2023){Lesur}, {Flock}, {Ercolano}, {Lin}, {Yang},
  {Barranco}, {Benitez-Llambay}, {Goodman}, {Johansen}, {Klahr}, {Laibe},
  {Lyra}, {Marcus}, {Nelson}, {Squire}, {Simon}, {Turner}, {Umurhan}, \&
  {Youdin}}]{Lesur2023}
{Lesur}, G., {Flock}, M., {Ercolano}, B., {et~al.} 2023, in Astronomical
  Society of the Pacific Conference Series, Vol. 534, Protostars and Planets
  VII, ed. S.~{Inutsuka}, Y.~{Aikawa}, T.~{Muto}, K.~{Tomida}, \& M.~{Tamura},
  465, \dodoi{10.48550/arXiv.2203.09821}

\bibitem[{{Lodato} \& {Price}(2010)}]{Lodato2010}
{Lodato}, G., \& {Price}, D.~J. 2010, \mnras, 405, 1212,
  \dodoi{10.1111/j.1365-2966.2010.16526.x}

\bibitem[{{Nealon} {et~al.}(2015){Nealon}, {Price}, \& {Nixon}}]{Nealon2015}
{Nealon}, R., {Price}, D.~J., \& {Nixon}, C.~J. 2015, \mnras, 448, 1526,
  \dodoi{10.1093/mnras/stv014}

\bibitem[{{Nixon} \& {King}(2016)}]{Nixon2016}
{Nixon}, C., \& {King}, A. 2016, in Lecture Notes in Physics, Berlin Springer
  Verlag, ed. F.~{Haardt}, V.~{Gorini}, U.~{Moschella}, A.~{Treves}, \&
  M.~{Colpi}, Vol. 905, 45, \dodoi{10.1007/978-3-319-19416-5_2}

\bibitem[{{Nixon} {et~al.}(2013{\natexlab{a}}){Nixon}, {King}, \&
  {Price}}]{Nixon2013}
{Nixon}, C., {King}, A., \& {Price}, D. 2013{\natexlab{a}}, \mnras, 434, 1946,
  \dodoi{10.1093/mnras/stt1136}

\bibitem[{{Nixon} {et~al.}(2013{\natexlab{b}}){Nixon}, {King}, \&
  {Price}}]{Nixon2013a}
---. 2013{\natexlab{b}}, \mnras, 434, 1946, \dodoi{10.1093/mnras/stt1136}

\bibitem[{{Nixon}(2012)}]{Nixon2012a}
{Nixon}, C.~J. 2012, \mnras, 423, 2597,
  \dodoi{10.1111/j.1365-2966.2012.21072.x}

\bibitem[{{Ogilvie}(1999)}]{Ogilvie1999}
{Ogilvie}, G.~I. 1999, \mnras, 304, 557,
  \dodoi{10.1046/j.1365-8711.1999.02340.x}

\bibitem[{{Papaloizou} \& {Pringle}(1983)}]{Paploizou1983}
{Papaloizou}, J.~C.~B., \& {Pringle}, J.~E. 1983, \mnras, 202, 1181,
  \dodoi{10.1093/mnras/202.4.1181}

\bibitem[{{Papaloizou} \& {Terquem}(1995)}]{papaloizou1995}
{Papaloizou}, J.~C.~B., \& {Terquem}, C. 1995, \mnras, 274, 987,
  \dodoi{10.1093/mnras/274.4.987}

\bibitem[{{Pelkonen} {et~al.}(2024){Pelkonen}, {Padoan}, {Juvela},
  {Haugb{\o}lle}, \& {Nordlund}}]{Pelkonen2024}
{Pelkonen}, V.-M., {Padoan}, P., {Juvela}, M., {Haugb{\o}lle}, T., \&
  {Nordlund}, {\r{A}}. 2024, arXiv e-prints, arXiv:2405.06520,
  \dodoi{10.48550/arXiv.2405.06520}

\bibitem[{{Pinte} {et~al.}(2016){Pinte}, {Dent}, {M{\'e}nard}, {Hales}, {Hill},
  {Cortes}, \& {de Gregorio-Monsalvo}}]{Pinte2016}
{Pinte}, C., {Dent}, W.~R.~F., {M{\'e}nard}, F., {et~al.} 2016, \apj, 816, 25,
  \dodoi{10.3847/0004-637X/816/1/25}

\bibitem[{{Price}(2012)}]{Price2012}
{Price}, D.~J. 2012, Journal of Computational Physics, 231, 759,
  \dodoi{10.1016/j.jcp.2010.12.011}

\bibitem[{{Price} \& {Federrath}(2010)}]{Price2010}
{Price}, D.~J., \& {Federrath}, C. 2010, \mnras, 406, 1659,
  \dodoi{10.1111/j.1365-2966.2010.16810.x}

\bibitem[{{Price} {et~al.}(2018){Price}, {Wurster}, {Tricco}, {Nixon},
  {Toupin}, {Pettitt}, {Chan}, {Mentiplay}, {Laibe}, {Glover}, {Dobbs},
  {Nealon}, {Liptai}, {Worpel}, {Bonnerot}, {Dipierro}, {Ballabio}, {Ragusa},
  {Federrath}, {Iaconi}, {Reichardt}, {Forgan}, {Hutchison}, {Constantino},
  {Ayliffe}, {Hirsh}, \& {Lodato}}]{Price2018}
{Price}, D.~J., {Wurster}, J., {Tricco}, T.~S., {et~al.} 2018, \pasa, 35, e031,
  \dodoi{10.1017/pasa.2018.25}

\bibitem[{{Rabago} {et~al.}(2024){Rabago}, {Zhu}, {Lubow}, \&
  {Martin}}]{Rabago2024}
{Rabago}, I., {Zhu}, Z., {Lubow}, S., \& {Martin}, R.~G. 2024, \mnras, 533,
  360, \dodoi{10.1093/mnras/stae1787}

\bibitem[{{Smallwood} {et~al.}(2021){Smallwood}, {Nealon}, {Chen}, {Martin},
  {Bi}, {Dong}, \& {Pinte}}]{Smallwood2021}
{Smallwood}, J.~L., {Nealon}, R., {Chen}, C., {et~al.} 2021, \mnras, 508, 392,
  \dodoi{10.1093/mnras/stab2624}

\bibitem[{{Teague} {et~al.}(2016){Teague}, {Guilloteau}, {Semenov}, {Henning},
  {Dutrey}, {Pi{\'e}tu}, {Birnstiel}, {Chapillon}, {Hollenbach}, \&
  {Gorti}}]{Teague2016}
{Teague}, R., {Guilloteau}, S., {Semenov}, D., {et~al.} 2016, \aap, 592, A49,
  \dodoi{10.1051/0004-6361/201628550}

\bibitem[{{Thies} {et~al.}(2011){Thies}, {Kroupa}, {Goodwin}, {Stamatellos}, \&
  {Whitworth}}]{Thies2011}
{Thies}, I., {Kroupa}, P., {Goodwin}, S.~P., {Stamatellos}, D., \& {Whitworth},
  A.~P. 2011, \mnras, 417, 1817, \dodoi{10.1111/j.1365-2966.2011.19390.x}

\bibitem[{{Young} {et~al.}(2023){Young}, {Stevenson}, {Nixon}, \&
  {Rice}}]{Young2023}
{Young}, A.~K., {Stevenson}, S., {Nixon}, C.~J., \& {Rice}, K. 2023, \mnras,
  525, 2616, \dodoi{10.1093/mnras/stad2451}

\end{thebibliography}
 \bibliographystyle{aasjournal}



\end{document}